\documentclass[11pt,twoside]{article}

\usepackage{asp2006}
\usepackage{epsf}
\usepackage{lscape}
\usepackage{graphicx}
\usepackage{amssymb, amsmath, amsbsy}
\newcommand{\ebv}{\ensuremath{E_{\rm B-V}}}

\newcommand{\ha}{H\ensuremath{\alpha}}
\newcommand{\hb}{H\ensuremath{\beta}}

\newcommand{\nii}{[N\,II]}
\newcommand{\sii}{[S\,II]}
\newcommand{\oiii}{[O\,III]}

\newcommand{\feii}{Fe\,II}

\begin{document}

\title{Exploring the Partially-obscured BLR and Partially-covered NLR} 

\author{K.~Zhang\altaffilmark{1}, T.-G.~Wang\altaffilmark{1},
X.-B.~Dong\altaffilmark{1}, H.-Y.~Zhou\altaffilmark{1},
H.-L.~Lu\altaffilmark{1}
 }

\altaffiltext{1}{Center for Astrophysics, University of Science and
Technology of China, Hefei, Anhui 230026, China;
zkdtc@mail.ustc.edu.cn, twang@ustc.edu.cn, xbdong@ustc.edu.cn}

\begin{abstract} 
We have conducted a series of investigations
on the geometry of the reddening material in AGNs,
which have important implications to the AGN unification and SMBH demography.
According to our statistics of partially obscured quasars,
we found that SMBHs in partially obscured type/phase (i.e., intermediate type)
are at least as abundant as normal quasars in the local Universe;
the reddening material in most objects are located in between the BLR and NLR.
According to our comparison of narrow lines in type 1 and 2 AGNs,
we found that for high-ionization or high-critical density narrow lines
(e.g. \oiii\ $\lambda 5007$, Balmer lines and \feii)
a significant fraction of the emission
arises from the inner dense part of the NLR;
this inner NLR is located very close to the central engine and
thus can be covered by the torus.
\end{abstract}

\section{Torus with respect to BLR and NLR}
The unification model, invoking a dusty torus located in between the
broad-line region (BLR) and narrow-line region (NLR)
of active galactic nuclei (AGNs),
is well supported for Seyfert galaxies.
However, there are still some important open questions:
Does the subtending angle of the torus vary with nuclear luminosity
(e.g., the `QSO 2' problem)?
What is the extent (geometry) of the torus
(e.g., to what extent does it cover and obscure the NLR,
and how does the blocking/obscuration depend on the orientation)?
What is its dust-to-gas ratio?
To address these questions, we have conducted a series of investigations
\textit{via} the partially obscured NLR and BLR,
and the partially covered NLR.\footnote{
By ``partially obscured'', we mean that the obscuration is moderate,
as indicated by the presence of large Balmer decrement;
by ``partially covered'', we mean that the inner part of the NLR
is covered by the torus, either totally or partially obscured. }

\section{Partially obscured BLR}

Partially obscured (i.e. intermediate type) AGNs are
ideal targets to study some of these questions.
Dong et al. (2005) have carried out a systematic search for partially obscured
quasars in the entire sample of the $z<0.3$ broad-line AGN in the SDSS EDR,
with the BLR extinction estimated from the large broad-line \ha/\hb\ ratios.
The use of broad-line Balmer decrement as an extinction indicator
is further justified in Dong et al. (2008);
they showed that the broad-line \ha/\hb\ ratios of blue Seyfert 1s and quasars
cluster around 3.06 with a tiny standard deviation of $\approx$0.03\,dex
and do not correlate with either the continuum slope, Eddington ratio, or luminosity.
According to their statistics, partially obscured quasars is at least as abundant
as normal quasars in the local Universe.
By a comparison of the Balmer decrements of the broad and narrow components,
they pointed out that the reddening of the NLR is much smaller than
that of the BLR in most of the partially obscured AGNs;
i.e., a dusty torus likely exists even in low-redshift quasars,
with large subtending angles (see Fig. 1).
Based on a larger homogeneous sample of about 9000 $z<0.35$ broad-line AGNs
culled from the SDSS DR4 according to our criteria (Dong et al., in preparation),
we have been deriving the internal \ebv\ distribution of the AGN BLR
in the local Universe, and hence to
obtain the fractions of obscured AGNs of various intrinsic luminosity
and of various degrees of extinction (Zhang et al., in preparation).
The significant merits of partially obscured AGNs are
to obtain a demography of SMBHs in the obscured type/phase,
to get information both of the AGN and of the host galaxy simultaneously
(thus to investigate the SMBH--bulge and starburst--AGN connections), etc.

Besides, partially obscured AGNs allow a reliable measurement of
the dust-to-gas ratio (e.g., Wang et al. 2005);
A dust-to-gas ratio as high as the Galactic value can be present in moderately
thick gas within a few parsecs of the AGN (e.g., Wang et al. 2009).
Moreover, by the monitoring of the variation of the dust extinction,
we can obtain information of the distribution, kinematics and even the origin
of the obscuring material (Wang et al. 2009).

\begin{figure}[tbp]
\centering
\includegraphics[width=1\textwidth]{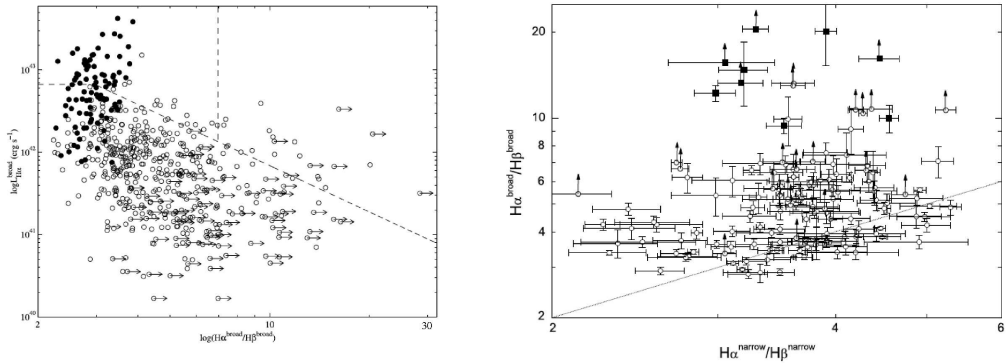}
\caption{
Left:
Luminosity of broad H$\alpha $ versus broad-line \ha/\hb\ ratio
for the broad-line AGN sample in the SDSS EDR.
Upper limits on the \ha/\hb\ ratio are tagged
with a right-pointing arrow.
Blue Seyfert 1s and QSOs are denoted by solid circles.
The inclined dash line corresponds to the estimated
M$_{g}^{nuc}=-$22$^{m}$.5 after internal extinction correction.
The partially obscured quasars scatter in the upper-right
region of the plot.
Right:
Broad-line H$\protect\alpha $/H$\protect\beta $ ratios versus narrow-line
H$\protect \alpha$/H$\protect\beta$ ratios of the intermediate type AGN in
the sample (Dong et al. 2005).
}
\end{figure}

\section{Partially covered NLR}
To investigate the extent of the torus,
we compare the narrow emission lines in type 1 and type 2 AGNs,
based on the entire AGN sample in the SDSS DR4 (Zhang et al. 2008).
We found that
(1) Seyfert 1 and Seyfert 2 galaxies have different distributions on the
\oiii/\hb\ versus \nii/\ha\ diagram (BPT diagram) for narrow lines;
(2) Among Seyfert 1 galaxies the distribution
varies with the extinction to broad lines;
and (3) The relationship between the \oiii\ and broad \ha\ luminosities
depends on the broad-line extinction in the way that
high-extinction objects have lower uncorrected \oiii\ luminosities
(see Fig. 2).

These results suggest that,
unlike low-ionization or low-critical density narrow lines such as \nii\ and \sii,
a significant fraction of the \oiii\ $\lambda 5007$
and Balmer-line emissions is blocked by the torus.
The inner edge of the dusty torus is known to be on scales
of parsecs; its extent, likely on scales of several tens of parsecs
(e.g., Schmitt et al. 2003, Jaffe et al. 2004).
Torus covers the inner dense part of the NLR.
This kind of partial covering causes the apparent anisotropy
(including the dependence on AGN type) of some narrow emission lines.
These narrow lines either are of high-ionization or high-critical density
(such as coronal lines and \oiii; see also Nagao et al. 2001) or
favor high-density and high-column density clouds (such as \feii;
Dong et al., 2009).

\begin{figure}
\centering
\includegraphics[width=1.\textwidth]{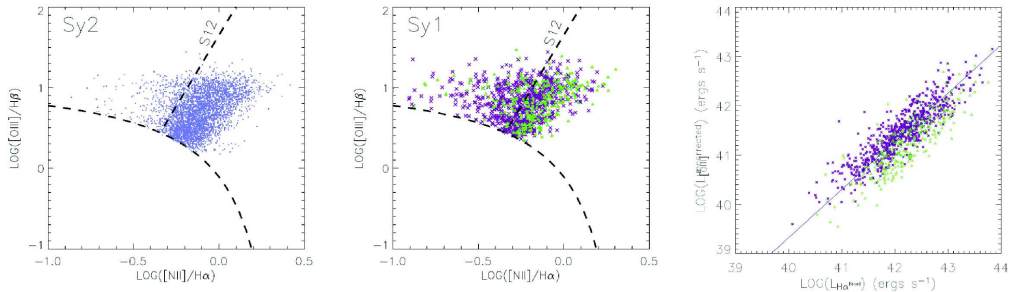}
\caption{
Left:
BPT diagram for Seyfert 2 galaxies.
The lower curve is the empirical line
separating AGNs from star-forming galaxies (Kewley et al 2006).
Most Seyfert 2s locate on the right side of the line S12.
Middle:
BPT diagram for Seyfert 1 galaxies.
Purple crosses represent objects with $E_{B-V}^{b}$\,$<$\,0.2
and green triangles those with $E_{B-V}^{b}$\,$\in$\,$[0.6,1]$.
Right:
The uncorrected luminosity of [O\,III]\,$\lambda 5007$
versus extinction corrected luminosity of broad H$\alpha$
for Seyfert 1 galaxies.
The blue line shows the best linear fit to the whole sample
(Zhang et al. 2008). }
\end{figure}


\end{document}